# BETATRON MOTION WITH COUPLING OF HORIZONTAL AND VERTICAL DEGREES OF FREEDOM[*]

V. A. Lebedev[♣]
*Fermi National Accelerator Laboratory, P.O. box 500, Batavia, IL60510*
S. A. Bogacz
*Thomas Jefferson National Accelerator Facility, 12000 Jefferson Avenue, Newport News, VA 23606*

Presntly, there are two most frequently used parameterezations of linear *x-y* coupled motion used in the accelerator physics. They are the Edwards-Teng and Mais-Ripken parameterizations. The article is devoted to an analysis of close relationship between the two representations, thus adding a clarity to their physical meaning. It also discusses the relationship between the eigen-vectors, the beta-functions, second order moments and the bilinear form representing the particle ellipsoid in the 4D phase space. Then, it consideres a further development of Mais-Ripken parameteresation where the particle motion is descrabed by 10 parameters: four beta-functions, four alpha-functions and two betatron phase advances. In comparison with Edwards-Teng parameterization the chosen parametrization has an advantage that it works equally well for analysis of coupled betatron motion in circular accelerators and in transfer lines. Considered relationship between second order moments, eigen-vectors and beta-functions can be useful in interpreting tracking results and experimental data. As an example, the developed formalizm is applied to the FNAL electron cooler and Derbenev's vertex-to-plane adapter.

## Introduction

In many applications analysis of coupled betatron motion is an important part of the machine design. The development of accelerator technology has stimulated additional interest in the subject in recent years. Initially betatron coupling was an undesired effect and efforts were made to suppress it. However, over recent two decades the betatron coupling has become an intrinsic part of many accelerator proposals[1-4]. Although many studies of the coupled motion have been performed over the last 40 years[5-14], in our opinion there is still no representation of coupled betatron motion that would be as elegant as the Courant-Snyder parametrization[15] for the one-dimensional case. Presently, two different basic representations are most frequently used. The first one was proposed by Edwards and Teng[5,6] and the second one by Mais and Ripken[7,8]. This article follows the steps of the second representation, where we limit the number of independent parameters to ten to parameterize a 4×4 symplectic transfer matrix. They are the four beta-functions, the four alpha-functions and the two betatron phase advances. The beta-functions have similar meaning to the Courant-Snyder parametrization, and the definition of alpha-functions coincides with the standard one in regions with zero longitudinal magnetic field, where they are equal to negative half-derivatives of the beta-functions. The article also reveals a close correspondence between the proposed parametrization and the Edwards-Teng parametrization, thus adding more clarity to their physical meaning.

The first section is mainly based on references [6], [8] and [16]. They describe the equations of motion, the notation and the basics of the theory developed in the 50's and the 60's. Section 2 sets relations between eigen-vectors, emittances and the particle 4D-ellipsoid in the phase space. Sections 3–5 develop the proposed representation and Section 6 shows its correspondence to the Edwards-Teng parametrization.

[*] Work supported by the US DOE under contract #DE-AC05-84ER40150 and by Fermi Research Alliance, LLC under Contract No. DE-AC02-07CH11359 with the U.S. Department of Energy
[♣] E-mail: val@jfnal.gov

## 1. Equations of Motion and Condition of Symplecticity

The two-dimensional linear motion of a particle in a focusing lattice structure can be described by the following set of equations:

$$x'' + (K_x^2 + k)x + \left(N - \frac{1}{2}R'\right)y - Ry' = 0 \quad ,$$
$$y'' + (K_y^2 - k)y + \left(N + \frac{1}{2}R'\right)x + Rx' = 0 \quad .$$
(1.1)

Here $x$ and $y$ are the horizontal and vertical particle displacements from the ideal orbit; the derivatives are calculated along the longitudinal coordinate $s$; $K_{x,y} = eB_{y,x}/Pc$; $k = eG/Pc$; $N = eG_s/Pc$; $R = eB_s/Pc$; $B_x$, $B_y$ and $B_s$ are the corresponding components of the magnetic field; $G$ is the normal component of the magnetic field gradient; and $G_s$ is the skew component of the magnetic field gradient (a quad tilted by +45 deg around the $s$ axis in the right-handed coordinate system).

The Hamiltonian[8] corresponding to Eq. (1.1) is

$$H = \frac{p_x^2 + p_y^2}{2} + \left(K_x^2 + k + \frac{R^2}{4}\right)\frac{x^2}{2} + \left(K_y^2 - k + \frac{R^2}{4}\right)\frac{y^2}{2} + Nxy + \frac{R}{2}(yp_x - xp_y) \quad , \quad (1.2)$$

and the corresponding canonical momenta are

$$p_x = x' - \frac{R}{2}y \quad ,$$
$$p_y = y' + \frac{R}{2}x \quad .$$
(1.3)

Rewriting Eq. (1.3) in matrix form we obtain the relation between the canonical, $\hat{\mathbf{x}}$, and the geometric coordinates, $\mathbf{x}$,

$$\hat{\mathbf{x}} = \mathbf{R}\mathbf{x} \quad , \quad (1.4)$$

where

$$\hat{\mathbf{x}} \equiv \begin{bmatrix} x \\ p_x \\ y \\ p_y \end{bmatrix} \quad , \quad \mathbf{x} \equiv \begin{bmatrix} x \\ \theta_x \\ y \\ \theta_y \end{bmatrix} \quad , \quad \mathbf{R} = \begin{bmatrix} 1 & 0 & 0 & 0 \\ 0 & 1 & -R/2 & 0 \\ 0 & 0 & 1 & 0 \\ R/2 & 0 & 0 & 1 \end{bmatrix} \quad , \quad (1.5)$$

$\theta_x = x'$ and $\theta_y = y'$. Here and below we put a cap above transfer matrices and vectors related to the canonical variables.

Introducing matrix $\mathbf{H}$,

$$\mathbf{H} = \begin{bmatrix} K_x^2 + k + \frac{R^2}{4} & 0 & N & -R/2 \\ 0 & 1 & R/2 & 0 \\ N & R/2 & K_y^2 - k + \frac{R^2}{4} & 0 \\ -R/2 & 0 & 0 & 1 \end{bmatrix} \quad , \quad (1.6)$$

one can rewrite Eqs. (1.1) and (1.2) in the matrix form:

$$\frac{d\hat{\mathbf{x}}}{ds} = \mathbf{U}\mathbf{H}\hat{\mathbf{x}} \quad , \tag{1.7}$$

$$H = \frac{1}{2}\hat{\mathbf{x}}^T \mathbf{H}\hat{\mathbf{x}} \quad , \tag{1.8}$$

where the unit symplectic matrix $\mathbf{U}$ is introduced as follows,

$$\mathbf{U} = \begin{bmatrix} 0 & 1 & 0 & 0 \\ -1 & 0 & 0 & 0 \\ 0 & 0 & 0 & 1 \\ 0 & 0 & -1 & 0 \end{bmatrix} \quad . \tag{1.9}$$

For any two solutions of Eq. (1.7), $\hat{\mathbf{x}}_1(s)$ and $\hat{\mathbf{x}}_2(s)$, one can write that

$$\frac{d}{ds}\left(\hat{\mathbf{x}}_1^T \mathbf{U}\hat{\mathbf{x}}_2\right) = \frac{d\hat{\mathbf{x}}_1^T}{ds}\mathbf{U}\hat{\mathbf{x}}_2 + \hat{\mathbf{x}}_1^T \mathbf{U}\frac{d\hat{\mathbf{x}}_2}{ds} = \hat{\mathbf{x}}_1^T \mathbf{H}^T \mathbf{U}^T \mathbf{U}\hat{\mathbf{x}}_2 + \hat{\mathbf{x}}_1^T \mathbf{U}\mathbf{U}\mathbf{H}\hat{\mathbf{x}}_2 = 0 \quad , \tag{1.10}$$

and, consequently,

$$\hat{\mathbf{x}}_1^T \mathbf{U}\hat{\mathbf{x}}_2 = \text{const} \quad , \tag{1.11}$$

where the following properties of the unit symplectic matrix were employed: $\mathbf{U}^T\mathbf{U} = \mathbf{I}$ and $\mathbf{U}\mathbf{U} = -\mathbf{I}$; and $\mathbf{I}$ is the identity matrix. The integral of motion of Eq. (1.11) is called the Lagrange invariant.

Let us introduce the transfer matrix from coordinate 0 to coordinate $s$, $\mathbf{x} = \mathbf{M}(0,s)\mathbf{x}_0$, and the corresponding transfer matrix for the canonical variables, $\hat{\mathbf{x}} = \hat{\mathbf{M}}(0,s)\hat{\mathbf{x}}_0$. Using Eq. (1.4) one finds that the matrices are bound up as following

$$\hat{\mathbf{M}}(0,s) = \mathbf{R}(s)\mathbf{M}(0,s)\mathbf{R}(0)^{-1} \quad . \tag{1.12}$$

Taking into account that the invariant of Eq. (1.11) does not change during motion, we can write that

$$\hat{\mathbf{x}}_1^T \mathbf{U}\hat{\mathbf{x}}_2 = \hat{\mathbf{x}}_1^T \hat{\mathbf{M}}(0,s)^T \mathbf{U}\hat{\mathbf{M}}(0,s)\hat{\mathbf{x}}_2 = \text{const} \quad . \tag{1.13}$$

As the above equation is satisfied for any $\hat{\mathbf{x}}_1$ and $\hat{\mathbf{x}}_2$ it yields

$$\hat{\mathbf{M}}(0,s)^T \mathbf{U}\hat{\mathbf{M}}(0,s) = \mathbf{U} \quad . \tag{1.14}$$

Eq. (1.14) expresses the symplecticity condition for particle motion. It is equivalent[16] to $n^2=16$ scalar equations, but taking into account that the matrix $\hat{\mathbf{M}}(0,s)^T \mathbf{U}\hat{\mathbf{M}}(0,s)$ is antisymmetric, only six $((n^2-n)/2 = 6)$ of these equations are independent. Consequently, only 10 of 16 elements of the transfer matrix are independent. Thus, the symplecticity condition imposes more severe limitations than the Liouville's theorem, which imposes only one condition, $\det(\mathbf{M})=1$, and leaves 15 independent parameters.

Consider a circular accelerator with the total transfer matrix $\hat{\mathbf{M}}$. The transfer matrix has four eigen-values, $\lambda_i$, and four corresponding eigen-vectors, $\hat{\mathbf{v}}_i$ ($i = 1, 2, 3, 4$),

$$\hat{\mathbf{M}}\hat{\mathbf{v}}_i = \lambda_i \hat{\mathbf{v}}_i \quad . \tag{1.15}$$

Below, we will consider the case of a stable betatron motion, meaning all four eigen-values are confined to a unit circle and none of them is equal to ±1. For any two eigen-vectors the symplecticity condition of Eq. (1.14) yields the identity

$$0 = \lambda_j \hat{\mathbf{v}}_j^T \mathbf{U}\left(\hat{\mathbf{M}}\hat{\mathbf{v}}_i - \lambda_i \hat{\mathbf{v}}_i\right) = \left(\hat{\mathbf{M}}\hat{\mathbf{v}}_j\right)^T \mathbf{U}\hat{\mathbf{M}}\hat{\mathbf{v}}_i - \lambda_j \hat{\mathbf{v}}_j^T \mathbf{U}\lambda_i \hat{\mathbf{v}}_i = \left(1 - \lambda_j \lambda_i\right)\hat{\mathbf{v}}_j^T \mathbf{U}\hat{\mathbf{v}}_i \quad , \tag{1.16}$$

which determines that the eigen-values always appear in two reciprocal pairs[8,16], and, consequently, the four eigen-values split into two complex conjugate pairs. We will denote them as $\lambda_1$, $\lambda_1^*$, $\lambda_2$ and $\lambda_2^*$, and the corresponding eigen-vectors as $\hat{\mathbf{v}}_1$, $\hat{\mathbf{v}}_1^*$, $\hat{\mathbf{v}}_2$ and $\hat{\mathbf{v}}_2^*$, where $^*$ denotes the complex conjugate value.

From Eq. (1.16) we obtain the following set of orthogonality conditions:

$$\hat{\mathbf{v}}_1^+ \mathbf{U} \hat{\mathbf{v}}_1 \neq 0,$$
$$\hat{\mathbf{v}}_2^+ \mathbf{U} \hat{\mathbf{v}}_2 \neq 0,$$
$$\hat{\mathbf{v}}_i^+ \mathbf{U} \hat{\mathbf{v}}_j = 0 \qquad \text{if } i \neq j,$$
$$\hat{\mathbf{v}}_i^T \mathbf{U} \hat{\mathbf{v}}_j = 0,$$

(1.17)

where $\hat{\mathbf{v}}^+ = \hat{\mathbf{v}}^{*T}$. The values in the two top lines of Eq.(1.17) are purely imaginary, indeed:

$$\left(\hat{\mathbf{v}}^+ \mathbf{U} \hat{\mathbf{v}}\right)^* = \left(\hat{\mathbf{v}}^+ \mathbf{U} \hat{\mathbf{v}}\right)^\dagger = \hat{\mathbf{v}}^+ \mathbf{U}^+ \hat{\mathbf{v}} = -\hat{\mathbf{v}}^+ \mathbf{U} \hat{\mathbf{v}}.$$

(1.18)

Therefore we normalize the eigen-vectors as follows:

$$\hat{\mathbf{v}}_1^+ \mathbf{U} \hat{\mathbf{v}}_1 = -2i, \qquad \hat{\mathbf{v}}_2^+ \mathbf{U} \hat{\mathbf{v}}_2 = -2i,$$
$$\hat{\mathbf{v}}_1^T \mathbf{U} \hat{\mathbf{v}}_1 = 0, \qquad \hat{\mathbf{v}}_2^T \mathbf{U} \hat{\mathbf{v}}_2 = 0,$$
$$\hat{\mathbf{v}}_2^T \mathbf{U} \hat{\mathbf{v}}_1 = 0, \qquad \hat{\mathbf{v}}_2^+ \mathbf{U} \hat{\mathbf{v}}_1 = 0.$$

(1.19)

Other combinations can be obtained by applying the transposition and/or the complex conjugation to Eqs. (1.19). Similarly as for the transfer matrix elements, there are only six independent real scalar equations among Eqs (1.19).

## 2. Relation between Eigen-vectors and Emittance Ellipsoid in 4D Phase Space

The turn-by-turn particle positions and angles (at the beginning of a lattice) can be represented as a linear combination of four independent solutions,

$$\hat{\mathbf{x}} = \text{Re}\left(A_1 e^{-i\psi_1} \hat{\mathbf{v}}_1 + A_2 e^{-i\psi_2} \hat{\mathbf{v}}_2\right)$$
$$= A_1\left(\hat{\mathbf{v}}_1' \cos\psi_1 + \hat{\mathbf{v}}_1'' \sin\psi_1\right) + A_2\left(\hat{\mathbf{v}}_2' \cos\psi_2 + \hat{\mathbf{v}}_1'' \sin\psi_2\right),$$

(2.1)

where four real parameters, $A_1$, $A_2$, $\psi_1$ and $\psi_2$, represent the betatron amplitudes and phases. The amplitudes remain constant in the course of betatron motion, while the phases change after each turn.

Let us introduce the following real matrix

$$\hat{\mathbf{V}} = \left[\hat{\mathbf{v}}_1', -\hat{\mathbf{v}}_1'', \hat{\mathbf{v}}_2', -\hat{\mathbf{v}}_2''\right].$$

(2.2)

This allows one to rewrite Eq. (2.1) in the compact form

$$\hat{\mathbf{x}} = \hat{\mathbf{V}} \mathbf{A} \boldsymbol{\xi}_A,$$

(2.3)

where the amplitude matrix $\mathbf{A}$ is

$$\mathbf{A} = \begin{bmatrix} A_1 & 0 & 0 & 0 \\ 0 & A_1 & 0 & 0 \\ 0 & 0 & A_2 & 0 \\ 0 & 0 & 0 & A_2 \end{bmatrix},$$

(2.4)

and

$$\xi_A = \begin{bmatrix} \cos\psi_1 \\ -\sin\psi_1 \\ \cos\psi_2 \\ -\sin\psi_2 \end{bmatrix} . \quad (2.5)$$

Applying the orthogonality conditions given by Eqs.(1.19), one can prove that matrix $\hat{\mathbf{V}}$ is a symplectic matrix. It can be seen explicitly as follows:

$$\hat{\mathbf{V}}^T \mathbf{U} \hat{\mathbf{V}} = \left[ \frac{\hat{\mathbf{v}}_1 + \hat{\mathbf{v}}_1^*}{2}, -\frac{\hat{\mathbf{v}}_1 - \hat{\mathbf{v}}_1^*}{2i}, \frac{\hat{\mathbf{v}}_2 + \hat{\mathbf{v}}_2^*}{2}, -\frac{\hat{\mathbf{v}}_2 - \hat{\mathbf{v}}_2^*}{2i} \right]^T \mathbf{U}$$
$$\left[ \frac{\hat{\mathbf{v}}_1 + \hat{\mathbf{v}}_1^*}{2}, -\frac{\hat{\mathbf{v}}_1 - \hat{\mathbf{v}}_1^*}{2i}, \frac{\hat{\mathbf{v}}_2 + \hat{\mathbf{v}}_2^*}{2}, -\frac{\hat{\mathbf{v}}_2 - \hat{\mathbf{v}}_2^*}{2i} \right] = \mathbf{U} . \quad (2.6)$$

Here we took into account that every matrix element in matrix $\hat{\mathbf{V}}^T \mathbf{U} \hat{\mathbf{V}}$ can be calculated using vector multiplication of Eqs. (1.19). Furthermore, the symplecticity of matrix $\hat{\mathbf{V}}$ yields the following useful expression for the inverse matrix, $\hat{\mathbf{V}}^{-1}$:

$$\hat{\mathbf{V}}^{-1} = -\mathbf{U}\hat{\mathbf{V}}^T \mathbf{U} , \quad (2.7)$$

where we took into account that $\mathbf{U}^T \mathbf{U} = \mathbf{I}$ and $\mathbf{U}^T = -\mathbf{U}$.

Let us consider an ensemble of particles, whose motion (at the beginning of lattice) is contained in a 4D ellipsoid. A 3D surface of this ellipsoid is determined by particles with extreme betatron amplitudes. For any of these particles, Eq. (2.3) describes the 2D-subspace of single-particle motion, which is a subspace of the 3D surface of the ellipsoid, described by the bilinear form

$$\hat{\mathbf{x}}^T \hat{\boldsymbol{\Xi}} \hat{\mathbf{x}} = 1 . \quad (2.8)$$

This ellipsoid confines the motion of all particles. To describe a 3D surface, in addition to parameters $\psi_1$ and $\psi_2$ of Eq. (2.5), we introduce the third parameter $\psi_3$ so that the vector $\boldsymbol{\xi}$ would describe a 3D sphere with a unit radius, according to the equation

$$(\boldsymbol{\xi}, \boldsymbol{\xi}) = 1 , \quad (2.9)$$

where

$$\boldsymbol{\xi} = \begin{bmatrix} \cos\psi_1 \cos\psi_3 \\ -\sin\psi_1 \cos\psi_3 \\ \cos\psi_2 \sin\psi_3 \\ -\sin\psi_2 \sin\psi_3 \end{bmatrix} . \quad (2.10)$$

Then, we can rewrite Eq. (2.3) in the following form,

$$\hat{\mathbf{x}} = \hat{\mathbf{V}} \mathbf{A} \boldsymbol{\xi} , \quad (2.11)$$

which describes a 3D subspace confining all particles of the beam. In other words we can consider that the amplitudes of the boundary particles are parameterized by $\psi_3$ ($A_1 \to A_1 \cos\psi_3$, $A_2 \to A_2 \cos\psi_3$.), so that we would obtain a 4D ellipsoid.

Expressing $\boldsymbol{\xi}$ from Eq. (2.11) and substituting it into Eq. (2.9), one obtains the quadratic form describing a 4D ellipsoid containing all particles:

$$\hat{\mathbf{x}}^T \left( \left(\hat{\mathbf{V}} \mathbf{A}\right)^{-1} \right)^T \left(\hat{\mathbf{V}} \mathbf{A}\right)^{-1} \hat{\mathbf{x}} = 1 . \quad (2.12)$$

Comparing Eqs. (2.8) and (2.12) and using Eq. (2.7), one can express the bilinear form, $\hat{\Xi}$, as follows:

$$\hat{\Xi} = \mathbf{U}\hat{\mathbf{V}}\hat{\Xi}'\hat{\mathbf{V}}^T\mathbf{U}^T \quad , \tag{2.13}$$

where $\hat{\Xi}' = \mathbf{A}^{-1}\mathbf{A}^{-1}$ is a diagonal matrix depending on two amplitudes $A_1$ and $A_2$, and we took into account that matrices $\mathbf{A}^{-1}$ and $\mathbf{U}$ commute.

To determine the beam emittance (volume of the occupied 4D phase-space) described by Eq. (2.8) we invert Eq. (2.13). That yields,

$$\hat{\Xi}' = \hat{\mathbf{V}}^T\hat{\Xi}\hat{\mathbf{V}} \quad . \tag{2.14}$$

As can be seen, a symplectic transform $\hat{\mathbf{V}}$ reduces matrix $\hat{\Xi}$ to its diagonal form. Then, in the new coordinate frame the 3D ellipsoid enclosing the total 4D phase-space of the beam can be described by the following equation:

$$\hat{\Xi}'_{11}x'^2 + \hat{\Xi}'_{22}p_x'^2 + \hat{\Xi}'_{33}y'^2 + \hat{\Xi}'_{44}p_y'^2 = 1 \quad . \tag{2.15}$$

It is natural to define the beam emittance as a product of the ellipsoid semi-axes (omitting the factor $\pi^2/2$ correcting for the real 4D volume of the ellipsoid) so that

$$\varepsilon_{4D} = \frac{1}{\sqrt{\hat{\Xi}'_{11}\hat{\Xi}'_{22}\hat{\Xi}'_{33}\hat{\Xi}'_{44}}} = \frac{1}{\sqrt{\det(\hat{\Xi}')}} \quad . \tag{2.16}$$

Calculation of the determinant using Eq. (2.14) yields,

$$\varepsilon_{4D} = \frac{1}{\sqrt{\det(\hat{\Xi}')}} = \frac{(A_1A_2)^2}{|\det(\hat{\mathbf{V}})|} = (A_1A_2)^2 \quad . \tag{2.17}$$

Here we took into account that $\det(\hat{\mathbf{V}}) = 1$, which is a direct consequence of matrix $\hat{\mathbf{V}}$ symplecticity. Thus, the squares of amplitudes $A_1$ and $A_2$ can be considered as 2D emittances $\varepsilon_1$ and $\varepsilon_2$ corresponding to the eigen-vectors $\hat{\mathbf{v}}_1$ and $\hat{\mathbf{v}}_2$. They coincide with the horizontal and vertical emittances of the uncoupled motion, and their product is equal to the total 4D emittance: $\varepsilon_1\varepsilon_2 = \varepsilon_{4D}$. Consequently, one can write matrix $\hat{\Xi}'$ as

$$\hat{\Xi}' = \begin{bmatrix} 1/\varepsilon_1 & 0 & 0 & 0 \\ 0 & 1/\varepsilon_1 & 0 & 0 \\ 0 & 0 & 1/\varepsilon_2 & 0 \\ 0 & 0 & 0 & 1/\varepsilon_2 \end{bmatrix} \quad . \tag{2.18}$$

Similarly to the one-dimensional case the particle ellipsoid shape, described by matrix $\hat{\Xi}$, determines the emittances $\varepsilon_1$ and $\varepsilon_2$, and the eigen-vectors $\hat{\mathbf{v}}_1$ and $\hat{\mathbf{v}}_2$. In this case the beam emittances are reciprocal to the roots of the following characteristic equation,

$$\det(\hat{\Xi} - i\lambda\mathbf{U}) = 0 \quad . \tag{2.19}$$

One can prove the above using Eqs. (2.13) as follows:

$$\det(\hat{\Xi} - i\lambda\mathbf{U}) = \det(\mathbf{U}\hat{\mathbf{V}}\hat{\Xi}'\hat{\mathbf{V}}^T\mathbf{U}^T - i\lambda\mathbf{U}) = \det(\hat{\Xi}' - i\lambda\mathbf{U}^T\hat{\mathbf{V}}^T\mathbf{U}\hat{\mathbf{V}}\mathbf{U}) =$$
$$\det(\hat{\Xi}' - i\lambda\mathbf{U}) = \left(\frac{1}{\varepsilon_1^2} - \lambda^2\right)\left(\frac{1}{\varepsilon_2^2} - \lambda^2\right) = 0 \quad . \tag{2.20}$$

Knowing the beam emittances and consequently $\hat{\Xi}'$, one can obtain from Eq. (2.13) a system of linear equations for matrix $\hat{\mathbf{V}}$,

$$\hat{\mathbf{\Xi}}\hat{\mathbf{V}}\mathbf{U} = \mathbf{U}\hat{\mathbf{V}}\hat{\mathbf{\Xi}}' \quad . \tag{2.21}$$

Multiplying the above equation by $\mathbf{u}_l$, one obtains two equations for the eigen-vectors:

$$\left(\hat{\mathbf{\Xi}} - \frac{i}{\varepsilon_l}\mathbf{U}\right)\hat{\mathbf{v}}_l = 0 \quad , \tag{2.22}$$

where $l = 1, 2$, and

$$\mathbf{u}_1 = \begin{bmatrix} 1 \\ -i \\ 0 \\ 0 \end{bmatrix} \quad , \quad \mathbf{u}_2 = \begin{bmatrix} 0 \\ 0 \\ 1 \\ -i \end{bmatrix} \quad . \tag{2.23}$$

We also took into account that $\hat{\mathbf{V}}\mathbf{u}_l = \hat{\mathbf{v}}_l$, $\mathbf{U}\mathbf{u}_l = -i\mathbf{u}_l$ and $\mathbf{\Xi}'\mathbf{u}_l = \frac{1}{\varepsilon_l}\mathbf{u}_l$.

Taking into account Eq. (2.8) a Gaussian distribution function for coupled beam motion can be written in the following form:

$$f(\hat{\mathbf{x}}) = \frac{1}{4\pi^2 \varepsilon_1 \varepsilon_2} \exp\left(-\frac{1}{2}\hat{\mathbf{x}}^T \hat{\mathbf{\Xi}} \hat{\mathbf{x}}\right) \quad . \tag{2.24}$$

Then, the second-order moments of the distribution function are

$$\hat{\Sigma}_{ij} \equiv \overline{\hat{x}_i \hat{x}_j} = \int \hat{x}_i \hat{x}_j f(\hat{\mathbf{x}}) d\hat{x}^4 = \frac{1}{4\pi^2 \varepsilon_1 \varepsilon_2} \int \hat{x}_i \hat{x}_j \exp\left(-\frac{1}{2}\hat{\mathbf{x}}^T \hat{\mathbf{\Xi}} \hat{\mathbf{x}}\right) d\hat{x}^4 \quad . \tag{2.25}$$

To carry out the integration one can perform a coordinate transform, $\hat{\mathbf{y}} = \hat{\mathbf{V}}^{-1}\hat{\mathbf{x}}$, which reduces matrix $\hat{\mathbf{\Xi}}$ to its diagonal form. After simple calculation one obtains the matrix of the second-order moments

$$\hat{\mathbf{\Sigma}} = \hat{\mathbf{V}} \begin{bmatrix} \varepsilon_1 & 0 & 0 & 0 \\ 0 & \varepsilon_1 & 0 & 0 \\ 0 & 0 & \varepsilon_2 & 0 \\ 0 & 0 & 0 & \varepsilon_2 \end{bmatrix} \hat{\mathbf{V}}^T \quad . \tag{2.26}$$

One can easily prove by direct substitution that matrix $\hat{\mathbf{\Sigma}}$ is the inverse of matrix $\hat{\mathbf{\Xi}}$. Consequently, a symplectic transform $\hat{\mathbf{V}}\mathbf{U}$ reduces matrix $\hat{\mathbf{\Sigma}}$ to its diagonal form. Applying a similar scheme as above for obtaining emittances and eigen-vectors from matrix $\hat{\mathbf{\Xi}}$, one finds that the beam emittances $\varepsilon_1$ and $\varepsilon_2$ can be computed from matrix $\hat{\mathbf{\Sigma}}$ as roots of its characteristic equation,

$$\det(\hat{\mathbf{\Sigma}}\mathbf{U} + i\lambda\mathbf{I}) = 0 \quad , \quad \varepsilon_l = \lambda_l \quad , \tag{2.27}$$

while the equations for the eigen-vectors are

$$(\hat{\mathbf{\Sigma}}\mathbf{U} + i\varepsilon_l\mathbf{I})\hat{\mathbf{v}}_l = 0 \quad . \tag{2.28}$$

It also follows from Eq. (2.26) that the total beam emittance is equal to

$$\varepsilon_{4D} = \varepsilon_1 \varepsilon_2 = \sqrt{\det(\hat{\mathbf{\Sigma}})} \quad . \tag{2.29}$$

Taking into account that the beam motion from point $s$ to point $s'$ results in the matrix $\mathbf{\Xi}$ transformation so that $\mathbf{\Xi}(s') = \mathbf{M}(s,s')^T \mathbf{\Xi}(s)\mathbf{M}(s,s')$, and using Eq. (2.19) and the motion simplicity one can easily prove that the emittances $\varepsilon_1$ and $\varepsilon_2$ are the motion invariants; *i.e.* there is no configuration of linear electric and magnetic fields which can change them. Taking into account that each emittance is bound to the corresponding betatron mode we will call them the mode emittances. If the beam line is build so that the motion is decoupled at some point then the mode emittances coincide with usual horizontal and vertical emittances.

Although the above analysis was carried out for the *x-y* coupled motion it can be directly applied to the coupling of any two degrees of freedom. If necessary it can be trivially extended to 3 or more degrees of freedom; so that the only change to be taking into account in Eqs. (2.12) – (2.29) is a larger dimension of the matrices and vectors.

## 3. Beta-functions for Coupled Motion

Employing the previously introduced notation, one can describe a single-particle phase-space trajectory along the beam orbit as

$$\hat{\mathbf{x}}(s) = \hat{\mathbf{M}}(0,s)\operatorname{Re}\left(\sqrt{\varepsilon_1}\hat{\mathbf{v}}_1 e^{-i\psi_1} + \sqrt{\varepsilon_2}\hat{\mathbf{v}}_2 e^{-i\psi_2}\right)$$
$$= \operatorname{Re}\left(\sqrt{\varepsilon_1}\hat{\mathbf{v}}_1(s)e^{-i(\psi_1+\mu_1(s))} + \sqrt{\varepsilon_2}\hat{\mathbf{v}}_2(s)e^{-i(\psi_2+\mu_2(s))}\right) \quad , \tag{3.1}$$

where the vectors $\hat{\mathbf{v}}_1(s) \equiv e^{i\mu_1(s)}\hat{\mathbf{M}}(0,s)\hat{\mathbf{v}}_1$ and $\hat{\mathbf{v}}_2(s) \equiv e^{i\mu_2(s)}\hat{\mathbf{M}}(0,s)\hat{\mathbf{v}}_2$ are the eigen-vectors of the matrix $\hat{\mathbf{M}}(0,s)\hat{\mathbf{M}}\hat{\mathbf{M}}(0,s)^{-1}$, $\psi_1$ and $\psi_2$ are the initial phases of betatron motion and $\hat{\mathbf{M}} = \hat{\mathbf{M}}(0,L)$ is the transfer matrix for the entire ring. The terms $e^{-i\mu_1(s)}$ and $e^{-i\mu_2(s)}$ are introduced to bring the eigen-vectors to the following standard form:

$$\hat{\mathbf{v}}_1(s) = \begin{bmatrix} \sqrt{\beta_{1x}(s)} \\ -\dfrac{iu_1(s)+\alpha_{1x}(s)}{\sqrt{\beta_{1x}(s)}} \\ \sqrt{\beta_{1y}(s)}e^{i\nu_1(s)} \\ -\dfrac{iu_2(s)+\alpha_{1y}(s)}{\sqrt{\beta_{1y}(s)}}e^{i\nu_1(s)} \end{bmatrix} \quad , \quad \hat{\mathbf{v}}_2(s) = \begin{bmatrix} \sqrt{\beta_{2x}(s)}e^{i\nu_2(s)} \\ -\dfrac{iu_3(s)+\alpha_{2x}(s)}{\sqrt{\beta_{2x}(s)}}e^{i\nu_2(s)} \\ \sqrt{\beta_{2y}(s)} \\ -\dfrac{iu_4(s)+\alpha_{2y}(s)}{\sqrt{\beta_{2y}(s)}} \end{bmatrix} \quad , \tag{3.2}$$

so that $\mu_1(s)$ and $\mu_2(s)$ would be the phase advances of betatron motion. Here $\beta_{1x}(s)$, $\beta_{1y}(s)$, $\beta_{2x}(s)$ and $\beta_{2y}(s)$ are the beta-functions; $\alpha_{1x}(s)$, $\alpha_{1y}(s)$, $\alpha_{2x}(s)$ and $\alpha_{2y}(s)$ are the alpha-functions which, as will be shown in the next section, coincide with the beta-functions' negative half-derivatives at regions with zero longitudinal magnetic field; and six real functions $u_1(s)$, $u_2(s)$, $u_3(s)$, $u_4(s)$, $\nu_1(s)$ and $\nu_2(s)$ are determined by the orthogonality conditions of Eq.(1.19). Below we will be omitting their dependence on *s* where it does not cause an ambiguity. Two eigen-vectors $\hat{\mathbf{v}}_1$ and $\hat{\mathbf{v}}_2$ were chosen out of two pairs of complex conjugate eigen-vectors by selecting $u_1$ and $u_4$ to be positive.

The first orthogonality condition of Eqs. (1.19),

$$\left(\hat{\mathbf{v}}_1^+ \mathbf{U}\hat{\mathbf{v}}_1\right) = -2i(u_1 + u_2) = -2i \quad , \tag{3.3}$$

yields $u_1 = 1 - u_2$, and similarly for the second eigen-vector, $u_4 = 1 - u_3$. The next two equations, $\hat{\mathbf{v}}_1^T \mathbf{U}\hat{\mathbf{v}}_1 = 0$ and $\hat{\mathbf{v}}_2^T \mathbf{U}\hat{\mathbf{v}}_2 = 0$, are identities.

Taking into account the above relations for $u_1$ and $u_4$, the remaining two non-trivial orthogonality conditions can be written as follows,

$$\left(\hat{\mathbf{v}}_2^+\mathbf{U}\hat{\mathbf{v}}_1\right) = -\left(\sqrt{\frac{\beta_{2x}}{\beta_{1x}}}[i(1-u_2)+\alpha_{1x}]+\sqrt{\frac{\beta_{1x}}{\beta_{2x}}}[iu_3-\alpha_{2x}]\right)e^{-iv_2}$$
$$-\left(\sqrt{\frac{\beta_{1y}}{\beta_{2y}}}[i(1-u_3)-\alpha_{2y}]+\sqrt{\frac{\beta_{2y}}{\beta_{1y}}}[iu_2+\alpha_{1y}]\right)e^{iv_1} = 0 \quad,$$

(3.4)

$$\left(\hat{\mathbf{v}}_2^T\mathbf{U}\hat{\mathbf{v}}_1\right) = -\left(\sqrt{\frac{\beta_{2x}}{\beta_{1x}}}[i(1-u_2)+\alpha_{1x}]-\sqrt{\frac{\beta_{1x}}{\beta_{2x}}}[iu_3+\alpha_{2x}]\right)e^{iv_2}$$
$$-\left(\sqrt{\frac{\beta_{1y}}{\beta_{2y}}}[i(u_3-1)-\alpha_{2y}]+\sqrt{\frac{\beta_{2y}}{\beta_{1y}}}[iu_2+\alpha_{1y}]\right)e^{iv_1} = 0 \quad.$$

(3.5)

Multiplying both terms in Eq.(3.4) and Eq.(3.5) by their complex conjugate values one obtains

$$A_x^2 + \left(\kappa_x(1-u_2)+\kappa_x^{-1}u_3\right)^2 = A_y^2 + \left(\kappa_y(1-u_3)+\kappa_y^{-1}u_2\right)^2 \quad,$$
$$A_x^2 + \left(\kappa_x(1-u_2)-\kappa_x^{-1}u_3\right)^2 = A_y^2 + \left(\kappa_y(1-u_3)-\kappa_y^{-1}u_2\right)^2 \quad,$$

(3.6)

where

$$A_x = \kappa_x\alpha_{1x} - \kappa_x^{-1}\alpha_{2x},$$
$$A_y = \kappa_y\alpha_{2y} - \kappa_y^{-1}\alpha_{1y},$$
$$\kappa_x = \sqrt{\frac{\beta_{2x}}{\beta_{1x}}}, \quad \kappa_y = \sqrt{\frac{\beta_{1y}}{\beta_{2y}}}.$$

(3.7)

Subtracting Eqs. (3.6) yields $u_2=u_3$. Substituting $u_2=u_3=u$ into the first equation of Eqs. (3.6) one obtains the following expression for $u$:

$$u = \frac{-\kappa_x^2\kappa_y^2 \pm \sqrt{\kappa_x^2\kappa_y^2\left(1+\frac{A_x^2-A_y^2}{\kappa_x^2-\kappa_y^2}\left(1-\kappa_x^2\kappa_y^2\right)\right)}}{1-\kappa_x^2\kappa_y^2} \quad.$$

(3.8)

By definition $u_k$ ($k = 1,\ldots 4$) are real functions[1] and $u_1$ and $u_4$ are positive. That sets a constraint for possible values of beta- and alpha-functions,

$$\frac{A_x^2-A_y^2}{\kappa_x^2-\kappa_y^2}\left(1-\kappa_x^2\kappa_y^2\right) \geq -1 \quad,$$

(3.9)

and a constraint on a value of $u$, $u \leq 1$.

Knowing $u$ makes it easy to find $v_1+v_2$ and $v_1-v_2$ from Eqs. (3.4) and (3.5):

$$e^{iv_+} \equiv e^{i(v_1+v_2)} = \frac{A_x + i\left(\kappa_x(1-u)+\kappa_x^{-1}u\right)}{A_y - i\left(\kappa_y(1-u)+\kappa_y^{-1}u\right)} \quad,$$
$$e^{iv_-} \equiv e^{i(v_1-v_2)} = \frac{A_x + i\left(\kappa_x(1-u)-\kappa_x^{-1}u\right)}{A_y + i\left(\kappa_y(1-u)-\kappa_y^{-1}u\right)} \quad,$$

(3.10)

---

[1] Eq. (3.8) also demonstrates that if beta- and alpha-functions are not correctly chosen, so that the value of the discriminant is negative, $u$ becomes imaginary, thus redetermining the alpha-functions.

and, consequently $\nu_1$ and $\nu_2$:

$$\nu_1 = \frac{1}{2}(\nu_+ + \nu_-) + \pi(n+m) \quad ,$$
$$\nu_2 = \frac{1}{2}(\nu_+ - \nu_-) + \pi(n-m) \quad .$$
(3.11)

Here $n$ and $m$ are arbitrary integers. Eq. (3.10) results in that $\nu_-$ and $\nu_+$ are determined modulo $2\pi$, which, consequently, yields that $\nu_1$ and $\nu_2$ are determined modulo $\pi$ (see Eq. (3.11)) resulting in additional solutions. Actually there are only two independent solutions for $\nu_1$ and $\nu_2$. The first one corresponds to the case when both $n$ and $m$ have the same parity, which is equivalent to $m+n = m-n = 0$. The second one corresponds to different parity of $m$ and $n$, which is equivalent to $m+n = m-n = 1$. Thus, in a general case, one has four independent solutions for $u$, $\nu_1$ and $\nu_2$ set by symplecticity conditions: two solutions for $u$ and two solutions for $\nu_1$ and $\nu_2$ for each $u$.

Below we will call thirteen functions, $\beta_{1x}$, $\beta_{1y}$, $\beta_{2x}$, $\beta_{2y}$, $\alpha_{1x}$, $\alpha_{1y}$, $\alpha_{2x}$, $\alpha_{2y}$, $u$, $\nu_1$, $\nu_2$, $\mu_1$ and $\mu_2$ the generalized Twiss functions. Only 10 of them are independent. Other three can be determined from the symplecticity conditions. Although for known eigen-vectors the Twiss parameters can be determined uniquely it is not the case if we know only beta-functions. In this case an application of symplecticity conditions leaves four independent solutions for the eigen-vectors. Two of them are related to the sign choice for $u$ in Eq. (3.8), and other two (for each choice of $u$) are related to uncertainty of $\nu_1$ and $\nu_2$ in Eq. (3.11). The later is related to the fact that the mirror reflection with respect to the $x$ or $y$ axis does not change $\beta$'s and $\alpha$'s but changes the relative signs for the $x$ and $y$ components of the eigen-vectors[2], with subsequent change of $\nu_1$ and $\nu_2$ by $\pi$. It is opposite to the case Edwards-Teng parameterization (see Section 6), where knowing eigen-vectors does not yield unique solution for the Twiss parameters but knowing Twiss parameters uniquely determines eigen-vectors.

Finally, we can express the eigen-vectors $\hat{\mathbf{v}}_1$ and $\hat{\mathbf{v}}_2$ in the following form:

$$\hat{\mathbf{v}}_1 = \begin{bmatrix} \sqrt{\beta_{1x}} \\ -\dfrac{i(1-u)+\alpha_{1x}}{\sqrt{\beta_{1x}}} \\ \sqrt{\beta_{1y}}e^{i\nu_1} \\ -\dfrac{iu+\alpha_{1y}}{\sqrt{\beta_{1y}}}e^{i\nu_1} \end{bmatrix} \quad , \quad \hat{\mathbf{v}}_2 = \begin{bmatrix} \sqrt{\beta_{2x}}e^{i\nu_2} \\ -\dfrac{iu+\alpha_{2x}}{\sqrt{\beta_{2x}}}e^{i\nu_2} \\ \sqrt{\beta_{2y}} \\ -\dfrac{i(1-u)+\alpha_{2y}}{\sqrt{\beta_{2y}}} \end{bmatrix} \quad .$$
(3.12)

That yields the following expression for matrix $\hat{\mathbf{V}}$ (see Eq.(2.2))

---

[2] It can also be achieved by change of the coupling sign (simultaneous sign change for gradients of all skew quads and magnetic fields of all solenoids), which does not change the beta-functions but does change the $\nu$-functions by $\pi$.

$$\hat{\mathbf{V}} = \begin{bmatrix} \sqrt{\beta_{1x}} & 0 & \sqrt{\beta_{2x}}\cos\nu_2 & -\sqrt{\beta_{2x}}\sin\nu_2 \\ -\dfrac{\alpha_{1x}}{\sqrt{\beta_{1x}}} & \dfrac{1-u}{\sqrt{\beta_{1x}}} & \dfrac{u\sin\nu_2 - \alpha_{2x}\cos\nu_2}{\sqrt{\beta_{2x}}} & \dfrac{u\cos\nu_2 + \alpha_{2x}\sin\nu_2}{\sqrt{\beta_{2x}}} \\ \sqrt{\beta_{1y}}\cos\nu_1 & -\sqrt{\beta_{1y}}\sin\nu_1 & \sqrt{\beta_{2y}} & 0 \\ \dfrac{u\sin\nu_1 - \alpha_{1y}\cos\nu_1}{\sqrt{\beta_{1y}}} & \dfrac{u\cos\nu_1 + \alpha_{1y}\sin\nu_1}{\sqrt{\beta_{1y}}} & -\dfrac{\alpha_{2y}}{\sqrt{\beta_{2y}}} & \dfrac{1-u}{\sqrt{\beta_{2y}}} \end{bmatrix}. \quad (3.13)$$

In the case of weak coupling one should normally choose $\hat{\mathbf{v}}_1$ as the eigen-vector, which mainly relates to the horizontal motion, and $\hat{\mathbf{v}}_2$ to the vertical motion. In the case of strong coupling the choice is arbitrary. As can be seen from Eq. (3.12), in determining beta- and alpha-functions, swapping two eigen-vectors causes the following redefinitions: $\beta_{1x} \leftrightarrow \beta_{2x}$, $\beta_{1y} \leftrightarrow \beta_{2y}$, $\alpha_{1x} \leftrightarrow \alpha_{2x}$, $\alpha_{1y} \leftrightarrow \alpha_{2y}$, $u \to 1-u$, $\nu_1 \to -\nu_2$ and $\nu_2 \to -\nu_1$. One can verify that Eqs. (3.8) and (3.10) satisfy the above transformations for $u$, $\nu_1$ and $\nu_2$.

To find the beam sizes one needs to remember that the amplitudes of beam motion related to the corresponding eigen-vectors are governed by Eqs. (2.10) and (2.11). Applying Eqs. (2.11), (3.1) and (3.12) one can parametrize the coordinates of the 4D ellipsoid interior:

$$\hat{\mathbf{x}}(\psi_1, \psi_2, \psi_3) = \mathrm{Re}\left(\sqrt{\varepsilon_1}\,\hat{\mathbf{v}}_1 \cos\psi_3\, e^{-i\psi_1} + \sqrt{\varepsilon_2}\,\hat{\mathbf{v}}_2 \sin\psi_3\, e^{-i\psi_2}\right). \quad (3.14)$$

The beam sizes (projections of 4D ellipsoid to the horizontal and vertical directions) are determined by the maximum of $x$ and $y$ variations in Eq.(3.14) and are equal to

$$a_x = \sqrt{\varepsilon_1 \beta_{1x} + \varepsilon_2 \beta_{2x}}, \\ a_y = \sqrt{\varepsilon_1 \beta_{1y} + \varepsilon_2 \beta_{2y}}. \quad (3.15)$$

Let us to write the equation describing the beam ellipsoid in the $x$-$y$ plane (the projection of the 4D-ellipsiod to the $x$-$y$ plane) in the following form,

$$\frac{x^2}{a_x^2} - \frac{2\tilde{\alpha} xy}{a_x a_y} + \frac{y^2}{a_y^2} = 1 - \tilde{\alpha}^2, \quad (3.16)$$

one can find the parameter $\tilde{\alpha}$ by determining at which $x$ coordinate the $y$ coordinate in Eq. (3.14) reaches the maximum. Comparing this result with the result following from Eq. (3.16) one obtains[8]:

$$\tilde{\alpha} = \frac{\sqrt{\beta_{1x}\beta_{1y}}\,\varepsilon_1 \cos\nu_1 + \sqrt{\beta_{2x}\beta_{2y}}\,\varepsilon_2 \cos\nu_2}{\sqrt{\varepsilon_1\beta_{1x} + \varepsilon_2\beta_{2x}}\sqrt{\varepsilon_1\beta_{1y} + \varepsilon_2\beta_{2y}}}. \quad (3.17)$$

Comparing Eqs. (3.15) and (3.17) to the second order moments presented in Appendix A one can see that the above beam sizes coincide with the rms beam sizes of the Gaussian distribution, and the parameter $\tilde{\alpha}$ can be also expressed as following $\tilde{\alpha} = \langle xy \rangle / \sqrt{\langle x^2 \rangle \langle y^2 \rangle}$.

## 4. Derivatives of the Tunes and Beta-Functions

Let us consider the relations between the beta- and alpha-functions. A differential trajectory displacement related to the first eigen-vector can be expressed as follows:

$$x(s+ds) = x(s) + x'(s)ds = x(s) + \left(p_x(s) + \frac{R}{2}y\right)ds =$$

$$\sqrt{\varepsilon_1}\,\text{Re}\left(\left(\sqrt{\beta_{1x}(s)} + \left[-\frac{i(1-u(s))+\alpha_{1x}(s)}{\sqrt{\beta_{1x}(s)}} + \frac{R}{2}\sqrt{\beta_{1y}(s)}\,e^{i\nu_1(s)}\right]ds\right)e^{-i(\mu_1(s)+\psi_1)}\right). \quad (4.1)$$

Alternatively, one can express particle position through the beta-functions at the new coordinate $s + ds$:

$$x(s+ds) = \text{Re}\left(\sqrt{\varepsilon_1 \beta_x(s+ds)}\,e^{-i(\mu_1(s+ds)+\psi)}\right) =$$

$$\sqrt{\varepsilon_1}\,\text{Re}\left(\left[\sqrt{\beta_{1x}(s)} + \frac{d\beta_{1x}}{2\sqrt{\beta_{1x}(s)}} - i\sqrt{\beta_{1x}(s)}\,d\mu\right]e^{-i(\mu_1(s)+\psi)}\right). \quad (4.2)$$

Comparing both the imaginary and real parts of Eqs. (4.1) and (4.2) one obtains:

$$\frac{d\beta_{1x}}{ds} = -2\alpha_{1x} + R\sqrt{\beta_{1x}\beta_{1y}}\cos\nu_1\;,$$

$$\frac{d\mu_1}{ds} = \frac{1-u}{\beta_{1x}} - \frac{R}{2}\sqrt{\frac{\beta_{1y}}{\beta_{1x}}}\sin\nu_1\;. \quad (4.3)$$

Similarly, one can write down equivalent expressions for the vertical displacement,

$$y(s+ds) = y(s) + y'(s)ds = y(s) + \left(p_y(s) - \frac{R}{2}x\right)ds =$$

$$\sqrt{\varepsilon_1}\,\text{Re}\left(\left(\sqrt{\beta_{1y}(s)}\,e^{i\nu_1(s)} - \left[\frac{iu(s)+\alpha_{1y}(s)}{\sqrt{\beta_{1y}(s)}}\,e^{i\nu_1(s)} + \frac{R}{2}\sqrt{\beta_{1x}(s)}\right]ds\right)e^{-i(\mu_1(s)+\psi_1)}\right), \quad (4.4)$$

and

$$y(s+ds) = \sqrt{\varepsilon_1}\,\text{Re}\left(\left[\sqrt{\beta_{1y}(s)} + \frac{d\beta_{1y}}{2\sqrt{\beta_{1y}(s)}} + i\sqrt{\beta_{1y}(s)}\,(d\nu_1 - d\mu_1)\right]e^{-i(\mu_1(s)+\psi-\nu_1(s))}\right), \quad (4.5)$$

which yields:

$$\frac{d\beta_{1y}}{ds} = -2\alpha_{1y} - R\sqrt{\beta_{1x}\beta_{1y}}\cos\nu_1\;,$$

$$\frac{d\mu_1}{ds} - \frac{d\nu_1}{ds} = \frac{u}{\beta_{1y}} + \frac{R}{2}\sqrt{\frac{\beta_{1x}}{\beta_{1y}}}\sin\nu_1\;. \quad (4.6)$$

Similar calculations carried out for the second eigen-vector yield,

$$\frac{d\beta_{2y}}{ds} = -2\alpha_{2y} - R\sqrt{\beta_{2x}\beta_{2y}}\cos\nu_2 \quad,$$

$$\frac{d\mu_2}{ds} = \frac{1-u}{\beta_{2y}} + \frac{R}{2}\sqrt{\frac{\beta_{2x}}{\beta_{2y}}}\sin\nu_2 \quad,$$

$$\frac{d\beta_{2x}}{ds} = -2\alpha_{2x} + R\sqrt{\beta_{2x}\beta_{2y}}\cos\nu_2 \quad,\qquad(4.7)$$

$$\frac{d\mu_2}{ds} - \frac{d\nu_2}{ds} = \frac{u}{\beta_{2x}} - \frac{R}{2}\sqrt{\frac{\beta_{2y}}{\beta_{2x}}}\sin\nu_2 \quad.$$

One can see that in the absence of longitudinal magnetic field the derivatives of the phase advances $d\mu_1/ds$ and $d\mu_2/ds$ are proportional to $(1-u)$ and are positive. That explains the selection rule for the eigen-vectors formulated at the beginning of Section 3 which requires $u_1$ and $u_4$ being positive ($u_1 = u_4 = 1 - u \geq 0$). Note that there is no a formal requirement for $d(\mu_1+\nu_1)/ds$ and $d(\mu_2+\nu_2)/ds$ being also positive and therefore $u$ can be negative[3], while in the most of practical cases it belongs to the [0,1] interval.

The relative contributions of $x$ and $y$ parts in the eigen-vector normalization equation, $\hat{\mathbf{v}}_l^+ \mathbf{U}\hat{\mathbf{v}}_l = -2i, l=1,2$, are proportional to $u$ or $1-u$. Therefore parameter $u$ can be considered as a coupling strength. In the absence of coupling the parameter $u$ is equal to 0 (or 1 if $x$ and $y$ vectors are swapped). Nevertheless, in the general case, an equality $u = 0$ does not mean an absence of coupling. As one can see from Eqs. (3.8) and (3.10) the condition $u=0$ requires $A_x = A_y$, and yields $e^{i\nu_+} = (A_x + i\kappa_x)/(A_x - i\kappa_y)$ and $e^{i\nu_-} = (A_x + i\kappa_x)/(A_y + i\kappa_y)$. These equations do not require auxiliary beta-functions $\beta_{1y}$ and $\beta_{2x}$ to be equal to zero, and, consequently, the condition $u = 0$ does not automatically mean absence of coupling. Although strictly speaking $u$ cannot be considered as a coupling parameter it reflects strength of the coupling and is a good value to characterize it in practice. In particular $u = \frac{1}{2}$ corresponds to 100% coupling when the motion for both eigen-vectors is equally distributed in both planes (see an example in Appendix B). It is also useful to note that $u$ does not change in an uncoupled transfer line. Actually, in the absence of coupling the $x$ and $y$ parts of the eigen-vector, $\hat{\mathbf{v}}_x$ and $\hat{\mathbf{v}}_y$, are independent and their normalization, $\hat{\mathbf{v}}_{x,y}^+ \mathbf{U}_2 \hat{\mathbf{v}}_{x,y} = \{u, 1-u\}$, does not change because the determinants of the corresponding 2×2 transfer matrices are equal to 1. Here $\mathbf{U}_2$ is the 2D unit symplectic matrix.

## 5. Representation of Transfer Matrix in Terms of Generalized Twiss Functions

One can derive a useful representation of the transfer matrix $\hat{\mathbf{M}}_{1,2} \equiv \hat{\mathbf{M}}(s_1, s_2)$ between two points of a transfer line in terms of the generalized Twiss functions. Using the definitions of eigen-vector and matrix $\hat{\mathbf{V}}$ (see Eqs.(3.1) and Eq.(2.2)) one can derive the following identity

---

[3] The Tevatron lattice is based on the detailed optics measurement and takes into account large coupling terms coming mainly from the skew-quadrupole components of the SC dipoles. If the coupling corrections are adjusted to minimize the tune split and, consequently, coupling the value of coupling parameter $u$ is normally varies in the range of about [-0.002, 0.04].

$$\hat{\mathbf{V}}_2 \mathbf{S} = \hat{\mathbf{M}}_{12} \hat{\mathbf{V}}_1 \quad . \tag{5.1}$$

Here $\hat{\mathbf{V}}_1$ and $\hat{\mathbf{V}}_2$ are the $\hat{\mathbf{V}}$ matrices given by Eq. (3.13) for the initial and final points. The matrix $\mathbf{S}$ is

$$\mathbf{S} = \begin{bmatrix} \cos\Delta\mu_1 & \sin\Delta\mu_1 & 0 & 0 \\ -\sin\Delta\mu_1 & \cos\Delta\mu_1 & 0 & 0 \\ 0 & 0 & \cos\Delta\mu_2 & \sin\Delta\mu_2 \\ 0 & 0 & -\sin\Delta\mu_2 & \cos\Delta\mu_2 \end{bmatrix} , \tag{5.2}$$

where $\Delta\mu_{1,2}$ are the betatron phase advances between points 1 and 2. Multiplying both sides of Eq.(5.1) by the inverse matrix, $\hat{\mathbf{V}}_1^{-1} = -\mathbf{U}\hat{\mathbf{V}}_1^T\mathbf{U}$, as given by Eq.(2.7), allows one to express the transfer matrix, $\hat{\mathbf{M}}_{12}$, in the form

$$\hat{\mathbf{M}}_{12} = -\hat{\mathbf{V}}_2 \mathbf{S} \hat{\mathbf{V}}_1^T \mathbf{U} \quad . \tag{5.3}$$

In the case of the one-turn transfer matrix $\hat{\mathbf{M}}$ the matrices $\hat{\mathbf{V}}_1$ and $\hat{\mathbf{V}}_2$ are equal and Eq. (5.3) simplifies. Explicit expressions of matrix $\hat{\mathbf{M}}$ as well as matrices $\hat{\mathbf{\Xi}}$ and $\hat{\mathbf{\Sigma}}$ are presented in Appendix A.

## 6. Edwards-Teng Parametrization

The Edwards-Teng parametrization[6] is based on a canonical transform $\tilde{\mathbf{R}}$ which reduces a 4×4 transfer matrix,

$$\hat{\mathbf{M}} = \begin{bmatrix} \mathbf{P} & \mathbf{p} \\ \mathbf{q} & \mathbf{Q} \end{bmatrix} , \tag{6.1}$$

to its normal modes form

$$\tilde{\mathbf{M}} = \tilde{\mathbf{R}}\hat{\mathbf{M}}\tilde{\mathbf{R}}^{-1} \quad , \tag{6.2}$$

where

$$\tilde{\mathbf{M}} = \begin{bmatrix} \mathbf{A} & \mathbf{0} \\ \mathbf{0} & \mathbf{B} \end{bmatrix} , \tag{6.3}$$

and $\mathbf{P}$, $\mathbf{p}$, $\mathbf{Q}$, $\mathbf{q}$, $\mathbf{A}$ and $\mathbf{B}$ are 2×2 matrices. Teng suggested parametrizing a symplectic matrix $\tilde{\mathbf{R}}$ as follows:

$$\tilde{\mathbf{R}} = \begin{bmatrix} \mathbf{E}\cos\phi & -\mathbf{D}^{-1}\sin\phi \\ \mathbf{D}\sin\phi & \mathbf{E}\cos\phi \end{bmatrix} , \tag{6.4}$$

where $\mathbf{E}$ is the unit 2×2 matrix, and $\mathbf{D}$ is a 2×2 symplectic matrix,

$$\mathbf{D} = \begin{bmatrix} a & b \\ c & d \end{bmatrix} , \tag{6.5}$$

so that $ad - bc = 1$. Thus, matrix $\tilde{\mathbf{R}}$ is parametrized by four parameters: *a*, *b*, *c* and *ϕ*. Matrix $\tilde{\mathbf{M}}$ describes the particle motion in new coordinates and can be parametrized by six Twiss parameters. Finally, one obtains ten parameters to fully describe the transfer matrix $\hat{\mathbf{M}}$. The six Twiss parameters $\beta_1$, $\alpha_1$, $\mu_1$, $\beta_2$, $\alpha_2$, and $\mu_2$ are so called the Twiss parameters of the decoupled motion. Edwards and Teng expressed them through the transfer matrix elements.

In the course of this section we will express them through the eigen-vectors. As will be seen below, this procedure reveals the close relation of the two representations and sheds additional light on the physical meaning of both parameter sets.

Expressing matrix $\hat{\mathbf{M}}$ through $\widetilde{\mathbf{M}}$ in Eq. (6.2) and substituting the result into Eq. (1.15), one obtains

$$\widetilde{\mathbf{R}}^{-1}\widetilde{\mathbf{M}}\widetilde{\mathbf{R}}\hat{\mathbf{v}}_i = \lambda_i \hat{\mathbf{v}}_i \quad . \tag{6.6}$$

Eq. (6.6) can be rewritten as

$$\widetilde{\mathbf{M}}\widetilde{\mathbf{v}}_i = \lambda_i \widetilde{\mathbf{v}}_i \quad , \tag{6.7}$$

where the vector

$$\widetilde{\mathbf{v}}_i = \widetilde{\mathbf{R}}\hat{\mathbf{v}}_i \tag{6.8}$$

is the eigen-vector of matrix $\widetilde{\mathbf{M}}$. To determine matrix $\widetilde{\mathbf{R}} \equiv \widetilde{\mathbf{R}}(s)$ we take into account that vectors $\widetilde{\mathbf{v}}_i$ represent decoupled motion; *i.e.*, the vector elements corresponding to another plane are equal to zero. Using the definitions of $\widetilde{\mathbf{R}}$, $\hat{\mathbf{v}}_i$ and expressing $\widetilde{\mathbf{v}}_i$ through the Twiss parameters of the decoupled motion, one can rewrite Eqs. (6.8) in the form:

$$\begin{bmatrix} \sqrt{\beta_1} \\ -\dfrac{i+\alpha_1}{\sqrt{\beta_1}} \\ 0 \\ 0 \end{bmatrix} = \begin{bmatrix} \cos\phi & 0 & -d\sin\phi & b\sin\phi \\ 0 & \cos\phi & c\sin\phi & -a\sin\phi \\ a\sin\phi & b\sin\phi & \cos\phi & 0 \\ c\sin\phi & d\sin\phi & 0 & \cos\phi \end{bmatrix} \begin{bmatrix} \sqrt{\beta_{1x}} \\ -\dfrac{i(1-u)+\alpha_{1x}}{\sqrt{\beta_{1x}}} \\ \sqrt{\beta_{1y}}e^{i\nu_1} \\ -\dfrac{iu+\alpha_{1y}}{\sqrt{\beta_{1y}}}e^{i\nu_1} \end{bmatrix} , \tag{6.9a}$$

$$\begin{bmatrix} 0 \\ 0 \\ \sqrt{\beta_2} \\ -\dfrac{i+\alpha_2}{\sqrt{\beta_2}} \end{bmatrix} = \begin{bmatrix} \cos\phi & 0 & -d\sin\phi & b\sin\phi \\ 0 & \cos\phi & c\sin\phi & -a\sin\phi \\ a\sin\phi & b\sin\phi & \cos\phi & 0 \\ c\sin\phi & d\sin\phi & 0 & \cos\phi \end{bmatrix} \begin{bmatrix} \sqrt{\beta_{2x}}e^{i\nu_2} \\ -\dfrac{iu+\alpha_{2x}}{\sqrt{\beta_{2x}}}e^{i\nu_2} \\ \sqrt{\beta_{2y}} \\ -\dfrac{i(1-u)+\alpha_{2y}}{\sqrt{\beta_{2y}}} \end{bmatrix} . \tag{6.9b}$$

Eqs. (6.9) represent eight scalar equations and they allow one to determine the parameters of matrix $\widetilde{\mathbf{R}}$ as well as the beta- and alpha-functions of the decoupled motion. Using the last two equations in Eq. (6.9a) and the first two equations in Eq. (6.9b), we obtain the following equations for matrix $\widetilde{\mathbf{R}}$ parameters:

$$\sqrt{\beta_{1x}}\, a_t - \frac{i(1-u)+\alpha_{1x}}{\sqrt{\beta_{1x}}} b_t + \sqrt{\beta_{1y}}\, e^{iv_1} = 0 \ ,$$

$$\sqrt{\beta_{1x}}\, c_t - \frac{i(1-u)+\alpha_{1x}}{\sqrt{\beta_{1x}}} d_t - \frac{iu+\alpha_{1y}}{\sqrt{\beta_{1y}}} e^{iv_1} = 0 \ ,$$

$$\sqrt{\beta_{2x}}\, e^{iv_2} - \sqrt{\beta_{2y}}\, d_t - \frac{i(1-u)+\alpha_{2y}}{\sqrt{\beta_{2y}}} b_t = 0 \ , \qquad (6.10)$$

$$-\frac{iu+\alpha_{2x}}{\sqrt{\beta_{2x}}} e^{iv_2} + \sqrt{\beta_{2y}}\, c_t + \frac{i(1-u)+\alpha_{2y}}{\sqrt{\beta_{2y}}} a_t = 0 \ .$$

Here the following notation was introduced: $a_t = a\tan\phi$, $b_t = b\tan\phi$, $c_t = c\tan\phi$ and $d_t = d\tan\phi$. Taking into account that $a_t$, $b_t$, $c_t$ and $d_t$ are real parameters, one can separate the real and imaginary parts in Eq. (6.10). That yields the following four solutions:

$$a_t = \sqrt{\frac{\beta_{2y}}{\beta_{2x}}} \frac{\alpha_{2x}\sin v_2 + u\cos v_2}{1-u} \ ,$$

$$b_t = \sqrt{\beta_{1x}\beta_{1y}}\, \frac{\sin v_1}{1-u} \ ,$$

$$c_t = \frac{\cos v_2 [\alpha_{2x}(1-u)-\alpha_{2y} u] - \sin v_2 [u(1-u)+\alpha_{2x}\alpha_{2y}]}{(1-u)\sqrt{\beta_{2x}\beta_{2y}}} \ , \qquad (6.11)$$

$$d_t = -\sqrt{\frac{\beta_{1x}}{\beta_{1y}}}\, \frac{u\cos v_1 + \alpha_{1y}\sin v_1}{1-u} \ ,$$

and four useful identities

$$\sqrt{\beta_{1x}\beta_{1y}}\sin v_1 = \sqrt{\beta_{2x}\beta_{2y}}\sin v_2 \ ,$$

$$\sqrt{\beta_{1x}\beta_{2y}}\,(\alpha_{2x}\sin v_2 + u\cos v_2) = \sqrt{\beta_{2x}\beta_{1y}}\,[\alpha_{1x}\sin v_1 - (1-u)\cos v_1] \ ,$$

$$\sqrt{\beta_{1x}\beta_{2y}}\,(\alpha_{1y}\sin v_1 + u\cos v_1) = \sqrt{\beta_{2x}\beta_{1y}}\,[\alpha_{2y}\sin v_2 - (1-u)\cos v_2] \ , \qquad (6.12)$$

$$\frac{(\alpha_{2x}\cos v_2 - u\sin v_2)(1-u) - (\alpha_{2x}\sin v_2 + u\cos v_2)\alpha_{2y}}{\sqrt{\beta_{2x}\beta_{2y}}} =$$

$$\frac{(\alpha_{1y}\cos v_1 - u\sin v_1)(1-u) - (\alpha_{1y}\sin v_1 + u\cos v_1)\alpha_{1x}}{\sqrt{\beta_{1x}\beta_{1y}}} \ .$$

The identities can be directly derived from the symplecticity of matrix $\hat{\mathbf{V}}$. Using Eq.(2.6) one immediately obtains that $\hat{\mathbf{V}}\mathbf{U}\hat{\mathbf{V}}^T\mathbf{U} = -\mathbf{I}$. Using the explicit definition of the matrix $\hat{\mathbf{V}}$ of Eq.(3.13) and performing matrix multiplication, after some algebra, one obtains these identities in the off-diagonal 2×2 block of the resulting matrix.

Using matrix **D** symplecticity and Eqs.(6.11), after simple algebra one obtains

$$\tan^2\phi = a_t d_t - b_t c_t = \frac{u}{1-u} \ . \qquad (6.13)$$

That finally yields:

$$\sin\phi = \pm\sqrt{u} \quad . \tag{6.14}$$

Now using the two first equations in Eq. (6.9a) and the two last equations in Eq. (6.9b), one obtains equations for the beta- and alpha-functions of the decoupled motion:

$$\begin{aligned}
\sqrt{\beta_1} &= \left(\sqrt{\beta_{1x}} - \sqrt{\beta_{1y}}e^{iv_1}d_t - \frac{iu+\alpha_{1y}}{\sqrt{\beta_{1y}}}e^{iv_1}b_t\right)\cos\phi \quad, \\
-\frac{i+\alpha_1}{\sqrt{\beta_1}} &= \left(-\frac{i(1-u)+\alpha_{1x}}{\sqrt{\beta_{1x}}} + \sqrt{\beta_{1y}}e^{iv_1}c_t + \frac{iu+\alpha_{1y}}{\sqrt{\beta_{1y}}}e^{iv_1}a_t\right)\cos\phi \quad, \\
\sqrt{\beta_2} &= \left(\sqrt{\beta_{2x}}e^{iv_2}a_t - \frac{iu+\alpha_{2x}}{\sqrt{\beta_{2x}}}e^{iv_2}b_t + \sqrt{\beta_{2y}}\right)\cos\phi \quad, \\
-\frac{i+\alpha_2}{\sqrt{\beta_2}} &= \left(\sqrt{\beta_{2x}}e^{iv_2}c_t - \frac{iu+\alpha_{2x}}{\sqrt{\beta_{2x}}}e^{iv_2}d_t - \frac{i(1-u)+\alpha_{2y}}{\sqrt{\beta_{2y}}}\right)\cos\phi \quad.
\end{aligned} \tag{6.15}$$

After lengthy calculation employing identities (6.12), one finally reduces the above equations to the simple form:

$$\begin{aligned}
\beta_1 &= \frac{\beta_{1x}}{1-u} \quad, \quad \alpha_1 = \frac{\alpha_{1x}}{1-u} \quad, \\
\beta_2 &= \frac{\beta_{2y}}{1-u} \quad, \quad \alpha_2 = \frac{\alpha_{2y}}{1-u} \quad.
\end{aligned} \tag{6.16}$$

As can be seen, although Eq. (6.14) yields four different values for angle $\phi$, other elements of matrix $\widetilde{\mathbf{R}}$ and the beta- and alpha-functions of the decoupled motion are uniquely related to the eigen-vectors and, consequently, to the generalized Twiss parameters. A problem appears if a value of $u$ is negative somewhere in the lattice. That results in $\phi$ being pure imaginary. The solution considered in Ref. [5] suggests a replacement of $\sin(\phi)$ and $\cos(\phi)$ by $\sinh(\phi)$ and $\cosh(\phi)$ with appropriate sign changes in the symplectic transforms of Eq. (6.9). It formally addresses the issue but still requires a redefinition of Eq.(6.9) symplectic transforms every time $u$ changes its sign.

The betatron motion in the normal modes representation can be written in the following form

$$\widetilde{\mathbf{x}}(s) = \widetilde{\mathbf{M}}(0,s)\widetilde{\mathbf{x}}(0) \quad, \tag{6.17}$$

where

$$\widetilde{\mathbf{M}}(0,s) = \widetilde{\mathbf{R}}(s)\hat{\mathbf{M}}(0,s)\widetilde{\mathbf{R}}^{-1}(0). \tag{6.18}$$

Edwards and Teng determined the phase advance of the betatron motion using a standard recipe for the decoupled motion:

$$\widetilde{\mathbf{v}}_i(s)e^{-i\mu_i(s)} = \widetilde{\mathbf{M}}(0,s)\widetilde{\mathbf{v}}_i(0) \quad. \tag{6.19}$$

Using the definition of matrix $\widetilde{\mathbf{M}}(0,s)$ of Eq. (6.18), we can rewrite Eq. (6.19) as

$$\hat{\mathbf{v}}_i(s)e^{-i\mu_i(s)} = \widetilde{\mathbf{R}}(s)^{-1}\widetilde{\mathbf{M}}(0,s)\widetilde{\mathbf{R}}(0)\hat{\mathbf{v}}_i(0) = \hat{\mathbf{M}}(0,s)\hat{\mathbf{v}}_i(0) \quad. \tag{6.20}$$

As can be seen, the obtained equation coincides with the definition of betatron phase advance of Section 4 (see Eq. (3.1) and below), thus proving that the betatron phase advances for both parametrizations are the same.

**Discussion**

This article introduces further development of the coupled betatron motion representation introduced in Refs. [6] and [7]. Our approach is based on a parametrization of the 4×4 symplectic transfer matrix by introducing ten functions: four beta-functions, four alpha-functions and two betatron phase advances, which we call the generalized Twiss functions. The beta-functions have similar meaning to the Courant-Snyder parametrization, and the definition of alpha-functions coincides with the definition for uncoupled motion at regions with zero longitudinal magnetic field, where they are equal to negative half-derivatives of the beta-functions. The approach is based on the parametrization of normalized eigen-vectors. Knowing the eigen-vectors, one can easily obtain the generalized betatron functions employing Eq.(3.12). Eqs. (3.8) and (3.10) allow one to perform the inverse operation of finding eigen-vectors from the generalized Twiss parameters. A useful representation of a transfer matrix in terms of the generalized Twiss functions is also introduced in Section 5.

A definition of 4D emittance is introduced for an ensemble of particles, whose motion is contained in a 4D ellipsoid. A 3D surface of this ellipsoid is determined by particles with extreme betatron amplitudes. Eqs. (2.8) and (A.2) determine the bilinear form $\hat{\Xi}$ describing this beam boundary. Consequently, the beam density distribution function can be written as

$$f(x, p_x, y, p_y) = A\delta(\hat{\mathbf{x}}^T \hat{\Xi} \hat{\mathbf{x}} - 1) \quad ,$$

in the case of KV-distribution, and as

$$f(x, p_x, y, p_y) = A\exp\left(-\frac{\hat{\mathbf{x}}^T \hat{\Xi} \hat{\mathbf{x}}}{2}\right) \quad ,$$

in the case of Gaussian distribution. The chosen normalization of the eigen-vectors, Eqs. (1.19), yields a simple relation between the beam emittances related to the eigen-vectors and total 4D emittance, $\varepsilon_{4D} = \varepsilon_1 \varepsilon_2$. Knowing the bilinear form $\hat{\Xi}$ or the matrix of second-order moments $\Sigma_{ij} \equiv \overline{\hat{x}_i \hat{x}_j}$, one can compute corresponding beam emittances, eigen-vectors and, consequently, generalized Twiss functions using Eqs. (2.19), (2.22) or Eqs. (2.27), (2.28). The mode emittances $\varepsilon_1$ and $\varepsilon_2$ are invariants of the motion.

A comparison of the developed parametrization with the Edwards-Teng parametrization provided additional insight for both parametrizations. First, it proved that the betatron motion phase advances for both parametrizations are equal; *i.e*,. the betatron phase advance for the Edwards-Teng representation is directly related to particle oscillations in the *x* or *y* plane, depending on which plane a particular eigen-vector is referenced to. Second, Edwards-Teng beta- and alpha-functions are simply related to the corresponding generalized beta- and alpha-functions: $\beta_i = \beta_{ix}/(1-u)$ , $\alpha_i = \alpha_{ix}/(1-u)$, where *u* is the coupling parameter directly related to the angle of Teng's rotation, $\sin^2 \phi = u$.

Unlike the Edwards-Teng parameterization the Mais-Ripken parameterization (as well as the parameterization developed in this article) allows one to obtain the unique solution for the generalized Twiss parameters from the known ring transfer matrix or the eigen-vectors. There are two linearly independent solutions in the case of Edwards-Teng parameterization. On the contrary, if one needs to determine the transfer matrix from the 10 Twiss parameters the Edwards-Teng parameterization yields the unique solution, while the parameterization developed in this article yields four solutions. To choose a unique solution one additionally needs to know which of two choices for $u$ and $\nu_1$ (or $\nu_2$) needs to be taken (see discussion after Eq. (3.11)).

The presented parametrization has been proven useful for both analytic and numerical analysis of coupled betatron motion in circular machines and transfer lines. Although we considered only $xy$-coupled motion in the article we would like to note that all results obtained in Section 2 are also applicable to three-dimensional particle motion. It is important to note that although the canonical coordinates were used throughout the article, this issue usually does not create complications in practical applications of the developed formalism because the canonical and geometric coordinates coincide at regions with zero longitudinal magnetic field. For example, the software developed by one of the authors for coupled-motion analysis always uses transfer matrices which start and end at points with zero longitudinal magnetic field, and thus, the canonical and geometric coordinates always coincide. Appendix B shows an example of analysis of how the strongly coupled motion for the Fermilab electron cooling project has been analyzed with the developed formalism.

*The authors are grateful to Y. Chao, G. Krafft, L. Harwood, S. Corneliussen and A. Burov for careful reading of the manuscript and useful suggestions for improving its clarity.*

## Appendix A. Explicit Expressions for Transfer Matrix, Bilinear Form and Matrix of Second Order Moments

Assuming one turn transformation and performing matrix multiplication in Eq.(5.3) one obtains the transfer matrix elements expressed through the generalized Twiss functions:

$$\hat{M}_{11} = (1-u)\cos\mu_1 + \alpha_{1x}\sin\mu_1 + u\cos\mu_2 + \alpha_{2x}\sin\mu_2 \quad, \tag{A.1.1}$$

$$\hat{M}_{12} = \beta_{1x}\sin\mu_1 + \beta_{2x}\sin\mu_2 \quad, \tag{A.1.2}$$

$$\hat{M}_{13} = \sqrt{\frac{\beta_{1x}}{\beta_{1y}}}\left[\alpha_{1y}\sin(\mu_1+\nu_1) + u\cos(\mu_1+\nu_1)\right] + \sqrt{\frac{\beta_{2x}}{\beta_{2y}}}\left[\alpha_{2y}\sin(\mu_2-\nu_2) + (1-u)\cos(\mu_2-\nu_2)\right] \quad, \tag{A.1.3}$$

$$\hat{M}_{14} = \sqrt{\beta_{1x}\beta_{1y}}\sin(\mu_1+\nu_1) + \sqrt{\beta_{2x}\beta_{2y}}\sin(\mu_2-\nu_2) \quad, \tag{A.1.4}$$

$$\hat{M}_{21} = -\frac{(1-u)^2+\alpha_{1x}^2}{\beta_{1x}}\sin\mu_1 - \frac{u^2+\alpha_{2x}^2}{\beta_{2x}}\sin\mu_2 \quad, \tag{A.1.5}$$

$$\hat{M}_{22} = (1-u)\cos\mu_1 + u\cos\mu_2 - \alpha_{1x}\sin\mu_1 - \alpha_{2x}\sin\mu_2 \quad, \tag{A.1.6}$$

$$\hat{M}_{23} = \frac{\left[(1-u)\alpha_{1y}-u\alpha_{1x}\right]\cos(\mu_1+\nu_1) - \left[\alpha_{1x}\alpha_{1y}+u(1-u)\right]\sin(\mu_1+\nu_1)}{\sqrt{\beta_{1x}\beta_{1y}}} +$$
$$\frac{\left[u\alpha_{2y}-(1-u)\alpha_{2x}\right]\cos(\mu_2-\nu_2) - \left[\alpha_{2x}\alpha_{2y}+u(1-u)\right]\sin(\mu_2-\nu_2)}{\sqrt{\beta_{2x}\beta_{2y}}} \quad, \tag{A.1.7}$$

$$\hat{M}_{24} = \sqrt{\frac{\beta_{1y}}{\beta_{1x}}}\left[(1-u)\cos(\mu_1+\nu_1) - \alpha_{1x}\sin(\mu_1+\nu_1)\right] + \sqrt{\frac{\beta_{2y}}{\beta_{2x}}}\left[u\cos(\mu_2-\nu_2) - \alpha_{2x}\sin(\mu_2-\nu_2)\right] \quad, \tag{A.1.8}$$

$$\hat{M}_{31} = \sqrt{\frac{\beta_{1y}}{\beta_{1x}}}\left[\alpha_{1x}\sin(\mu_1-\nu_1) + (1-u)\cos(\mu_1-\nu_1)\right] + \sqrt{\frac{\beta_{2y}}{\beta_{2x}}}\left[\alpha_{2x}\sin(\mu_2+\nu_2) + u\cos(\mu_2+\nu_2)\right] \quad, \tag{A.1.9}$$

$$\hat{M}_{32} = \sqrt{\beta_{1x}\beta_{1y}}\sin(\mu_1-\nu_1) + \sqrt{\beta_{2x}\beta_{2y}}\sin(\mu_2+\nu_2) \quad, \tag{A.1.10}$$

$$\hat{M}_{33} = u\cos\mu_1 + (1-u)\cos\mu_2 + \alpha_{2y}\sin\mu_2 + \alpha_{1y}\sin\mu_1 , \tag{A.1.11}$$

$$\hat{M}_{34} = \beta_{1y}\sin\mu_1 + \beta_{2y}\sin\mu_2 , \tag{A.1.12}$$

$$\hat{M}_{41} = \frac{[\alpha_{1x}u - (1-u)\alpha_{1y}]\cos(\mu_1 - \nu_1) - [\alpha_{1x}\alpha_{1y} + u(1-u)]\sin(\mu_1 - \nu_1)}{\sqrt{\beta_{1x}\beta_{1y}}} + \frac{[(1-u)\alpha_{2x} - u\alpha_{2y}]\cos(\mu_2 + \nu_2) - [\alpha_{2x}\alpha_{2y} + u(1-u)]\sin(\mu_2 + \nu_2)}{\sqrt{\beta_{2x}\beta_{2y}}} , \tag{A.1.13}$$

$$\hat{M}_{42} = \sqrt{\frac{\beta_{1x}}{\beta_{1y}}}[u\cos(\mu_1 - \nu_1) - \alpha_{1y}\sin(\mu_1 - \nu_1)] + \sqrt{\frac{\beta_{2x}}{\beta_{2y}}}[(1-u)\cos(\mu_2 + \nu_2) - \alpha_{2y}\sin(\mu_2 + \nu_2)] , \tag{A.1.14}$$

$$\hat{M}_{43} = -\frac{u^2 + \alpha_{1y}^2}{\beta_{1y}}\sin\mu_1 - \frac{(1-u)^2 + \alpha_{2y}^2}{\beta_{2y}}\sin\mu_2 , \tag{A.1.15}$$

$$\hat{M}_{44} = u\cos\mu_1 + (1-u)\cos\mu_2 - \alpha_{1y}\sin\mu_1 - \alpha_{2y}\sin\mu_2 . \tag{A.1.16}$$

Similarly, using Eq. (2.13), one can express elements of the bilinear form describing the particle ellipsoid in 4D space:

$$\hat{\Xi}_{11} = \frac{(1-u)^2 + \alpha_{1x}^2}{\varepsilon_1 \beta_{1x}} + \frac{u^2 + \alpha_{2x}^2}{\varepsilon_2 \beta_{2x}} , \tag{A.2.1}$$

$$\hat{\Xi}_{22} = \frac{\beta_{1x}}{\varepsilon_1} + \frac{\beta_{2x}}{\varepsilon_2} , \tag{A.2.2}$$

$$\hat{\Xi}_{33} = \frac{u^2 + \alpha_{1y}^2}{\varepsilon_1 \beta_{1y}} + \frac{(1-u)^2 + \alpha_{2y}^2}{\varepsilon_2 \beta_{2y}} , \tag{A.2.3}$$

$$\hat{\Xi}_{44} = \frac{\beta_{1y}}{\varepsilon_1} + \frac{\beta_{2y}}{\varepsilon_2} , \tag{A.2.4}$$

$$\hat{\Xi}_{12} = \hat{\Xi}_{21} = \frac{\alpha_{1x}}{\varepsilon_1} + \frac{\alpha_{2x}}{\varepsilon_2} , \tag{A.2.5}$$

$$\hat{\Xi}_{34} = \hat{\Xi}_{43} = \frac{\alpha_{1y}}{\varepsilon_1} + \frac{\alpha_{2y}}{\varepsilon_2} , \tag{A.2.6}$$

$$\hat{\Xi}_{13} = \hat{\Xi}_{31} = \frac{[\alpha_{1x}\alpha_{1y} + u(1-u)]\cos\nu_1 + [\alpha_{1y}(1-u) - \alpha_{1x}u]\sin\nu_1}{\varepsilon_1\sqrt{\beta_{1x}\beta_{1y}}} + \frac{[\alpha_{2x}\alpha_{2y} + u(1-u)]\cos\nu_2 + [\alpha_{2x}(1-u) - \alpha_{2y}u]\sin\nu_2}{\varepsilon_2\sqrt{\beta_{2x}\beta_{2y}}} , \tag{A.2.7}$$

$$\hat{\Xi}_{14} = \hat{\Xi}_{41} = \sqrt{\frac{\beta_{1y}}{\beta_{1x}}}\frac{\alpha_{1x}\cos\nu_1 + (1-u)\sin\nu_1}{\varepsilon_1} + \sqrt{\frac{\beta_{2y}}{\beta_{2x}}}\frac{\alpha_{2x}\cos\nu_2 - u\sin\nu_2}{\varepsilon_2} , \tag{A.2.8}$$

$$\hat{\Xi}_{23} = \hat{\Xi}_{32} = \sqrt{\frac{\beta_{1x}}{\beta_{1y}}}\frac{\alpha_{1y}\cos\nu_1 - u\sin\nu_1}{\varepsilon_1} + \sqrt{\frac{\beta_{2x}}{\beta_{2y}}}\frac{\alpha_{2y}\cos\nu_2 + (1-u)\sin\nu_2}{\varepsilon_2} , \tag{A.2.9}$$

$$\hat{\Xi}_{24} = \hat{\Xi}_{42} = \frac{\sqrt{\beta_{1x}\beta_{1y}}\cos\nu_1}{\varepsilon_1} + \frac{\sqrt{\beta_{2x}\beta_{2y}}\cos\nu_2}{\varepsilon_2} \qquad . \tag{A.2.10}$$

Finally, using Eq. (2.26), one can express elements of the second-order moments:

$$\hat{\Sigma}_{11} \equiv \langle x^2 \rangle = \varepsilon_1 \beta_{1x} + \varepsilon_2 \beta_{2x} \quad , \tag{A.3.1}$$

$$\hat{\Sigma}_{12} \equiv \langle x p_x \rangle = \hat{\Sigma}_{21} = -\varepsilon_1 \alpha_{1x} - \varepsilon_2 \alpha_{2x} \quad , \tag{A.3.2}$$

$$\hat{\Sigma}_{22} \equiv \langle p_x^2 \rangle = \varepsilon_1 \frac{(1-u)^2 + \alpha_{1x}^2}{\beta_{1x}} + \varepsilon_2 \frac{u^2 + \alpha_{2x}^2}{\beta_{2x}} \quad , \tag{A.3.3}$$

$$\hat{\Sigma}_{33} \equiv \langle y^2 \rangle = \varepsilon_1 \beta_{1y} + \varepsilon_2 \beta_{2y} \quad , \tag{A.3.4}$$

$$\hat{\Sigma}_{34} \equiv \langle y p_y \rangle = \hat{\Sigma}_{43} = -\varepsilon_1 \alpha_{1y} - \varepsilon_2 \alpha_{2y} \quad , \tag{A.3.5}$$

$$\hat{\Sigma}_{44} \equiv \langle p_y^2 \rangle = \varepsilon_1 \frac{u^2 + \alpha_{1y}^2}{\beta_{1y}} + \varepsilon_2 \frac{(1-u)^2 + \alpha_{2y}^2}{\beta_{2y}} \quad , \tag{A.3.6}$$

$$\hat{\Sigma}_{13} \equiv \langle xy \rangle = \hat{\Sigma}_{31} = \varepsilon_1 \sqrt{\beta_{1x}\beta_{1y}} \cos\nu_1 + \varepsilon_2 \sqrt{\beta_{2x}\beta_{2y}} \cos\nu_2 \quad , \tag{A.3.7}$$

$$\hat{\Sigma}_{14} \equiv \langle x p_y \rangle = \hat{\Sigma}_{41} = \varepsilon_1 \sqrt{\frac{\beta_{1x}}{\beta_{1y}}} \left( u \sin\nu_1 - \alpha_{1y} \cos\nu_1 \right) - \varepsilon_2 \sqrt{\frac{\beta_{2x}}{\beta_{2y}}} \left( (1-u) \sin\nu_2 + \alpha_{2y} \cos\nu_2 \right) \quad , \tag{A.3.8}$$

$$\hat{\Sigma}_{23} \equiv \langle y p_x \rangle = \hat{\Sigma}_{32} = -\varepsilon_1 \sqrt{\frac{\beta_{1y}}{\beta_{1x}}} \left( (1-u) \sin\nu_1 + \alpha_{1x} \cos\nu_1 \right) + \varepsilon_2 \sqrt{\frac{\beta_{2y}}{\beta_{2x}}} \left( u \sin\nu_2 - \alpha_{2x} \cos\nu_2 \right) \quad , \tag{A.3.9}$$

$$\hat{\Sigma}_{24} \equiv \langle p_x p_y \rangle = \hat{\Sigma}_{42} = \varepsilon_1 \frac{(\alpha_{1y}(1-u) - \alpha_{1x}u)\sin\nu_1 + (u(1-u) + \alpha_{1x}\alpha_{1y})\cos\nu_1}{\sqrt{\beta_{1x}\beta_{1y}}} + \varepsilon_2 \frac{(\alpha_{2x}(1-u) - \alpha_{2y}u)\sin\nu_2 + (u(1-u) + \alpha_{2x}\alpha_{2y})\cos\nu_2}{\sqrt{\beta_{2x}\beta_{2y}}} \quad . \tag{A.3.10}$$

**Appendix B. Generalized Twiss Functions for Axisymmetric Distribution Function**

To increase Tevatron luminosity, Fermilab developed a high-energy electron cooling device for the cooling of antiprotons[2]. Because of the high energy of the electron beam (~4 MeV), it is impractical to use the standard beam transport used in electron cooling devices where the beam moves in the longitudinal magnetic field the entire way from the electron gun to the collector. Nevertheless the longitudinal magnetic field is still used for beam focusing in the cooling section to cancel the beam defocusing due to the electron beam space charge, and more importantly to prevent collective instability in the electron beam. To neutralize the rotational motion of particles in the cooling section, the beam is produced in the electron gun immersed in the longitudinal magnetic field. Consequently, the beam transport is quite sophisticated, with a large number of bends and focusing elements. Taking into account that the space-charge effects are comparatively small everywhere except the gun and the collector, the developed formalism has been used for analysis of the main part of beam transport. In this section we consider how to find the generalized Twiss parameters and the mode emittances at the beginning of transport line.

At the exit of the electrostatic accelerator the electron beam distribution is axially symmetric, and before the beam leaves the magnetic field its distribution function is uncoupled and can be described by the bilinear form

$$\Xi_B = \frac{1}{\varepsilon_T} \begin{bmatrix} \gamma_0 & \alpha_0 & 0 & 0 \\ \alpha_0 & \beta_0 & 0 & 0 \\ 0 & 0 & \gamma_0 & \alpha_0 \\ 0 & 0 & \alpha_0 & \beta_0 \end{bmatrix}, \quad (B.1)$$

where $\varepsilon_T = r_c \sqrt{mkT_c}/P_0$ is the thermal emittance of the beam, $r_c$ is the cathode radius, $T_c$ is the cathode temperature, $P_0$ and $m$ are the particle momentum and mass, $\beta_0 = a^2/\varepsilon_T$, $\alpha_0 = -\sqrt{\beta_0/\varepsilon_T}(da/ds)$ and $\gamma_0 = (1+\alpha_0^2)/\beta_0$ are the initial Twiss functions, and $a$ is the beam radius at the electrostatic accelerator exit. We imply here that $a$ and $r_c$ can be different due to adiabatic beam expansion in the solenoid. Consequently, magnetic fields at the cathode and the solenoid exit are related: $B_c r_c^2 = B a^2$. After exiting from the magnetic field an electron acquires the angular momentum proportional to its radius, and the distribution can be characterized by the bilinear form:

$$\Xi_{in} = \Phi^T \Xi_B \Phi = \frac{1}{\varepsilon_T} \begin{bmatrix} \gamma_0 + \Phi^2 \beta_0 & \alpha_0 & 0 & -\Phi\beta_0 \\ \alpha_0 & \beta_0 & \Phi\beta_0 & 0 \\ 0 & \Phi\beta_0 & \gamma_0 + \Phi^2 \beta_0 & \alpha_0 \\ -\Phi\beta_0 & 0 & \alpha_0 & \beta_0 \end{bmatrix}, \quad (B.2)$$

where

$$\Phi = \begin{bmatrix} 1 & 0 & 0 & 0 \\ 0 & 1 & \Phi & 0 \\ 0 & 0 & 1 & 0 \\ -\Phi & 0 & 0 & 1 \end{bmatrix}, \quad (B.3)$$

$\Phi = eB/2P_0 c$ is the rotational focusing strength of the solenoid edge, and $B$ is the magnetic field at solenoid exit.

To choose initial values for generalized Twiss functions[4] we use the axial symmetry of the electron distribution function. This implies that the horizontal and vertical alpha- and beta-functions are equal and $u=1/2$. Thus, we obtain for the eigen-vectors:

$$\hat{\mathbf{v}}_1 = \begin{bmatrix} \sqrt{\beta} \\ -\dfrac{i+2\alpha}{2\sqrt{\beta}} \\ \sqrt{\beta} e^{i\nu_1} \\ -\dfrac{i+2\alpha}{2\sqrt{\beta}} e^{i\nu_1} \end{bmatrix}, \quad \hat{\mathbf{v}}_2 = \begin{bmatrix} \sqrt{\beta} e^{i\nu_2} \\ -\dfrac{i+2\alpha}{2\sqrt{\beta}} e^{i\nu_2} \\ \sqrt{\beta} \\ -\dfrac{i+2\alpha}{2\sqrt{\beta}} \end{bmatrix}. \quad (B.4)$$

In this case the coefficients of Eq. (3.7) are

$$\kappa_x = \kappa_y = 1 \quad \text{and} \quad A_x = A_y = 0 , \tag{B.5}$$

which creates uncertainty in Eqs. (3.8) and (3.10) for $u$, $v_1$ and $v_2$. To avoid this uncertainty we will use primarily Eqs. (3.4) and (3.5). Substituting Eqs.(B.4) into Eq.(3.4) yields

$$e^{-iv_1} + e^{iv_2} = 0 , \tag{B.6}$$

while for Eq.(3.5) it yields an identity. The solution of Eq.(B.6) is $v_1 = -v_2 + 2\pi(n + 1/2)$. As one can see there are an unlimited number of solutions for $v_1$ and $v_2$. We will choose a solution reflecting the eigen-vectors symmetry: $v_1 = v_2 = \pi/2$. Then, the matrix $\hat{\mathbf{V}}$ is equal to:

$$\hat{\mathbf{V}} = \begin{bmatrix} \sqrt{\beta} & 0 & 0 & -\sqrt{\beta} \\ -\dfrac{\alpha}{\sqrt{\beta}} & \dfrac{1}{2\sqrt{\beta}} & \dfrac{1}{2\sqrt{\beta}} & \dfrac{\alpha}{\sqrt{\beta}} \\ 0 & -\sqrt{\beta} & \sqrt{\beta} & 0 \\ \dfrac{1}{2\sqrt{\beta}} & \dfrac{\alpha}{\sqrt{\beta}} & -\dfrac{\alpha}{\sqrt{\beta}} & \dfrac{1}{2\sqrt{\beta}} \end{bmatrix} . \tag{B.7}$$

---

[4] We could use Eqs. (2.19) and (2.22) for computing the emittances and eigen-vectors and, consequently, the generalized Twiss functions, but it would require significantly more complicated calculations than for the procedure described below .

Using Eq. (2.13) (compare also with Eqs. (A.2)) we obtain the bilinear form,

$$\Xi = \begin{bmatrix} \dfrac{1+4\alpha^2}{4\beta}\left(\dfrac{1}{\varepsilon_1}+\dfrac{1}{\varepsilon_2}\right) & \alpha\left(\dfrac{1}{\varepsilon_1}+\dfrac{1}{\varepsilon_2}\right) & 0 & \dfrac{1}{2}\left(\dfrac{1}{\varepsilon_1}-\dfrac{1}{\varepsilon_2}\right) \\ \alpha\left(\dfrac{1}{\varepsilon_1}+\dfrac{1}{\varepsilon_2}\right) & \beta\left(\dfrac{1}{\varepsilon_1}+\dfrac{1}{\varepsilon_2}\right) & -\dfrac{1}{2}\left(\dfrac{1}{\varepsilon_1}-\dfrac{1}{\varepsilon_2}\right) & 0 \\ 0 & -\dfrac{1}{2}\left(\dfrac{1}{\varepsilon_1}-\dfrac{1}{\varepsilon_2}\right) & \dfrac{1+4\alpha^2}{4\beta}\left(\dfrac{1}{\varepsilon_1}+\dfrac{1}{\varepsilon_2}\right) & \alpha\left(\dfrac{1}{\varepsilon_1}+\dfrac{1}{\varepsilon_2}\right) \\ \dfrac{1}{2}\left(\dfrac{1}{\varepsilon_1}-\dfrac{1}{\varepsilon_2}\right) & 0 & \alpha\left(\dfrac{1}{\varepsilon_1}+\dfrac{1}{\varepsilon_2}\right) & \beta\left(\dfrac{1}{\varepsilon_1}+\dfrac{1}{\varepsilon_2}\right) \end{bmatrix} . \tag{B.8}$$

Comparing Eqs. (B.2) and (B.8), one can express generalized Twiss functions through the Twiss parameters of the beam distribution function in the magnetic field:

$$\beta = \frac{\beta_0}{2\sqrt{1+\Phi^2\beta_0^2}} , \quad \alpha = \frac{\alpha_0}{2\sqrt{1+\Phi^2\beta_0^2}} ,$$
$$\varepsilon_1 = \frac{\varepsilon_T}{\sqrt{1+\Phi^2\beta_0^2} - \Phi\beta_0} , \quad \varepsilon_2 = \frac{\varepsilon_T}{\sqrt{1+\Phi^2\beta_0^2} + \Phi\beta_0} . \tag{B.9}$$

One can see that $\varepsilon_1\varepsilon_2 = \varepsilon_T^2$, which verifies the conclusions of Section 2. The last two equations demonstrate that after exiting the magnetic field the beam distribution is characterized by two quite different emittances. In the case of FNAL cooler where $\Phi\beta_0 \gg 1$ it results in one emittance to be much larger another one. The first emittance is determined by the angular momentum excited by the solenoid edge field, $\varepsilon_1 = eB_c r_c^2/(P_0 c)$ and grows with the field. While the second emittance is determined by the cathode temperature, $\varepsilon_1 = mkT_c c/(eB_c P_0)$, and decreases with field increase.

The developed formalism presents also a simple way to describe the vertex-to-plane transform suggested by Derbenev[1]. As it was presented above, the eigen-vectors of Eq.(B.4) represent the vertex distribution function for $v_1 = v_2 = \pi/2$, while for $v_1 = 0$ and $v_2 = \pi$ they correspond to the uncoupled motion, in which $x$ and $y$ coordinates were rotated by $\pi/4$. The transform from one to another set of the eigen-vectors can be performed with a matrix representing a decoupled motion with betatron phase advances for the $x$ and $y$ planes different by $\pi/2$. In the case of unequal emittances $\varepsilon_1$ and $\varepsilon_2$ the initially axial-symmetric beam is transformed to an elliptic beam tilted by $\pi/4$. If the focusing system is rotated by $\pi/4$, the final elliptical beam is also rotated by the same angle due to the axial symmetry of the initial distribution. The final beam has an uncoupled distribution function with the emittances $\varepsilon_1$ and $\varepsilon_2$ corresponding to the vertical and horizontal emittances.